\documentclass[a4paper,11pt]{article}
\pdfoutput=1 

\usepackage{jcappub} 

\usepackage[T1]{fontenc} 
\DeclareMathOperator\arctanh{arctanh}

\title{\boldmath Non-minimally coupled varying constants quantum cosmologies}

\author[a,b]{Adam Balcerzak}

\affiliation[a]{Institute of Physics, University of Szczecin,  \\Wielkopolska 15, 70-451 Szczecin,  Poland}
\affiliation[b]{Copernicus Center for Interdisciplinary Studies,   \\S{\l }awkowska 17, 31-016 Krak\'ow, Poland}

\emailAdd{abalcerz@wmf.univ.szczecin.pl}

\abstract{We consider gravity theory with varying speed of light and varying gravitational constant. Both constants are represented by non-minimally coupled  scalar fields. We examine the cosmological  evolution in  the near curvature singularity regime. We find that at the curvature singularity the speed of light goes to infinity while the gravitational constant vanishes. This corresponds to the Newton's Mechanics limit represented by one of the vertex of the Bronshtein-Zelmanov-Okun cube~\cite{Duff,Okun}. The cosmological evolution includes both the pre-big-bang and post-big-bang phases separated by the curvature singularity. We also investigate the quantum counterpart of the considered theory and find the probability of transition of the universe from the collapsing pre-big-bang phase to the expanding post-big-bang phase.}

\begin{document}
\maketitle
\flushbottom

\section{Introduction}
\label{sec:intro}

Among the most popular theories admitting  varying  physical constants are those involving the variable fine structure constant $\alpha$, the speed of light $c$, and the gravitational constant $G$. A lot of important motivation for such theories have been provided on both theoretical and experimental basis~\cite{Uzan}.

The most important  framework for the variability of the gravitational constant is provided by the scalar-tensor theories, where the effective gravitational constant is determined by the local value of the scalar field~\cite{Brans}.

Theories with variable speed of light can generally be divided into two groups. The first group contains theories  that assume the dependence of the speed of light on the space-time coordinates. Such an assumption unavoidably leads to violation of the Lorentz invariance~\cite{Albrecht,Barrow1,Clayton,Drummond,Clayton2,Magueijo1}. In the second group there are theories where the speed of light depends on the energy scale~\cite{Rainbow}.
 The most radical example of a theory violating Lorentz invariance assumes the existence of a preferred frame usually identified with the cosmological frame~\cite{Albrecht,Barrow1}.  In these theories  the dynamics of the gravitational field,  matter fields and the scalar field representing the speed of light is described by gravity theory  with minimal coupling in the preferred frame. It is also assumed that the scalar field which determines the value of the speed of light  does not contribute to the energy momentum tensor. Consequently, in the preferred frame the gravitational field is described by the standard Einstein equations with the speed of light being merely replaced by the above mentioned scalar field.
The bimetric  VSL theory~\cite{Clayton,Drummond,Clayton2} is another example of a theory that brakes  Lorentz invariance. In  this framework the  two separate  metrics are introduced. One of them describes the geometry of  space-time while the other  couples to the matter. As a consequence, in such a theory the speed of gravitational waves may differ from the speed of massless particles.
Another interesting subclass of theories with varying speed of light exploits a concept of  local Lorentz invariance which is applicable when the speed of light is allowed to vary~\cite{Magueijo1}.
The  second class contains theories assuming a modified dispersion relation,  wherein the modification becomes important for the energy scales  comparable with the Planck scale. As a  consequence in the considered  model the group velocity of light becomes dependent on the energy scale. Originally, such theories  assumed  the modified dispersion relation to be fulfilled  in the preferred frame and hence they explicitly  violated  the Lorentz invariance~\cite{Amelino,Ellis,Ellis2}. As it was shown in~\cite{Amelino2,Kowalski},  it is possible to preserve  the Lorentz invariance (its non-linear form),  while still assuming modified dispersion relation.

In refs.~\cite{Harko,Szydlowski,Yurov}  the tunneling from nothing to a de Sitter phase within the quantum cosmology with varying physical $c$ and (or) $G$ constants was discussed.
In ref.~\cite{Leszczynska} the theory of gravity with varying $G$ and $c$, with the minimal coupling in the cosmological frame was considered. Neither $c$ nor $G$ did contributed to the energy momentum tensor. Consequently, the cosmological equations were of the form of the standard  Friedmann equations with  both constants replaced by arbitrary  time dependent functions. By using the quantized theory described by the one-dimensional Wheeler-De Witt equation the probability of tunneling of the universe from nothing was derived.
A similar approach was adopted in ref.~\cite{Garattini} where it was shown that the cosmological constant can be considered as an eigenvalue of an appropriate Sturm-Liouville problem of a varying speed of light theory.
We think it would be  advisable to consider  a theory where both varying $c$ and $G$  represent additional degrees of freedom. This could be achieved by replacing c and G  with two  separate scalar fields.

Our paper is organized as follows. In Sec. II we formulate a theory with varying $c$ and $G$ where both constants are represented by non-minimally coupled scalar fields and find the classical solution near the curvature singularity. We show that the cosmological evolution is equivalent to the process of the one-dimensional scattering of the point particle on the exponential potential barrier. In Sec.  III we solve the Wheeler-De Witt equation for the quantum counterpart of the considered theory and find that the quantum behavior is equivalent to the one-dimensional quantum stationary scattering of the point particle on the exponential potential barrier. By calculating the probability of scattering we find the probability of the transition from the pre-big-bang collapsing phase to the post-big-bang expanding phase. In Sec. IV we give our conclusions.

\section{Non-minimally coupled classical varying $c$ and $G$ theory}
We consider a theory which  admits  varying speed of light $c$ and varying gravitational constant $G$. This is achieved by connecting  both constants $c$ and $G$ with the two  scalar fields $\phi$ and $\psi$. Additionally, we include separate kinetic terms for each scalar field. Such a theory can be represented by the following action:
\begin{equation}
\label{action}
S=\int \sqrt{-g}  \left(\frac{e^{\phi}}{e^{\psi}}\right) \left[R+\Lambda + \omega (\partial_\mu \phi \partial^\mu \phi + \partial_\mu \psi \partial^\mu \psi)\right] d^4x~,
\end{equation}
where $\omega$ is a constant parameter and the relations between varying  $c(x^\mu)$ and $G(x^\mu)$ and the scalar fields $\phi(x^\mu)$ and $\psi(x^\mu)$ are given by  $c^3=e^{\phi}$ and $G=e^\psi$. By the application of  the following field transformations:
\begin{equation}
\label{fred}
\begin{split}
\phi &= \frac{\beta}{\sqrt{2\omega}}+\frac{1}{2} \ln \delta \,,
\\
\psi &= \frac{\beta}{\sqrt{2\omega}}-\frac{1}{2} \ln \delta \,.
\end{split}
\end{equation}
the action \eqref{action} can be rewritten in the form of the Brans-Dicke action:
\begin{equation}
\label{actionBD}
S=\int \sqrt{-g}\left[ \delta (R+\Lambda) +\frac{\omega}{2}\frac{\partial_\mu \delta \partial^\mu \delta}{\delta} + \delta \partial_\mu \beta \partial^\mu \beta\right]d^4x~,
\end{equation}
where the constant parameter $\omega$  can now be identified with the Brans-Dicke parameter.

We consider a Friedmann  universe described by the FLRW metric:
\begin{equation}
\label{metric}
ds^2=-N(t)^2 (dx^0)^2+a(t)^2\left(\frac{dr^2}{1-k r^2}+r^2 d\Omega^2\right)~,
\end{equation}
where $N(t)$ is the lapse function, $a(t)$ the scale factor and $k=0,\pm 1$.
By inserting the metric \eqref{metric} into the action  \eqref{actionBD} and then  integrating by parts we get:
\begin{equation}
\label{action sym}
S = \frac{3 V_0}{8 \pi} \int dx^0 \left(-\frac{a^2}{N} a' \delta' - \frac{\delta}{N} a a'^2 + k a \delta N  + \Lambda \delta a^3 N -\frac{\omega}{2} \frac{a^3}{N} \frac{\delta'^2}{\delta}-\frac{a^3}{N}\delta \beta'^2 \right)~.
\end{equation}
where $(~)'=\frac{\partial}{\partial x^0}$. By choosing the lapse function as $N=a^3 \delta$ and further  by application of the fields transformations given by:
\begin{equation}
\label{fred2}
\begin{split}
X &= \ln(a \sqrt{\delta}) \,,
\\
Y &= \frac{1}{2A} \ln \delta \,,
\end{split}
\end{equation}
where $A=\frac{1}{\sqrt{1-2\omega}}$, one arrives in the following form of the action:
\begin{equation}
S=\frac{3 V_0}{8 \pi} \int dx^0 \left[ -X'^2 + Y'^2 + e^{4X}\left(k+\Lambda e^{2(X-AY)} \right)-\beta'^2\right]~.
\end{equation}
From now on, we restrict our considerations to the spatially flat Friedmann universes and put $k=0$. Further fields redefinition  $I=AY-3X$,  $J=3Y-AX$ and $B=\sqrt{\tilde{V}_0}\beta$ gives:
\begin{equation}
\label{action final}
S= \int dx^0 \left[m (J'^2-I'^2) + \bar{\Lambda} e^{-2I}-B'^2\right]~,
\end{equation}
where we have defined the following constant parameters:
\begin{equation*}
\label{eq:x}
\begin{split}
\tilde{V}_0 = \frac{3 V_0}{8\pi} \,,
\qquad
m = \frac{\tilde{V}_0}{9-A^2} \,,
\qquad
\bar{\Lambda} = \tilde{V}_0\Lambda \,.
\end{split}
\end{equation*}
Equivalently, our theory can be described by the following Hamiltonian:
\begin{equation}
\label{ham}
H=\frac{1}{4}\left[\frac{1}{m}\left(\pi_J^2- \pi_I^2\right)- \pi_B^2\right] -\bar{\Lambda}e^{-2I}~,
\end{equation}
where $\pi_J=2mJ'$, $\pi_I=-2mI'$, and $\pi_B=-2B'$ are  the canonically conjugated momenta. The form of the hamiltonian \eqref{ham} clearly shows that $\pi_J$ and $\pi_B$ are conserved. Consequently, the classical evolution near the singularity can be reduced to the problem of motion of the point particle  in the one-dimensional exponential potential.
The associated Hamilton equations read as:
\begin{equation}
\label{ham eq}
\begin{split}
B' &= \frac{\partial H}{\partial \pi_B}=-\frac{\pi_B}{2}\,,
\qquad
\pi_B' = -\frac{\partial H}{\partial B}=0\,,
\\
I' &= \frac{\partial H}{\partial \pi_I}=-\frac{\pi_I}{2m}\,,
\qquad
\pi_I' = -\frac{\partial H}{\partial I}=-2 \bar{\Lambda} e^{-2I} \,,
\\
J'&=\frac{\partial H}{\partial \pi_J} = \frac{\pi_J}{2m}\,,
\qquad
\pi_J' = - \frac{\partial H}{\partial J}=0 \,.
\end{split}
\end{equation}
The solution of the Hamilton equations \eqref{ham eq} is given by:
\begin{equation}
\label{ham sol}
\begin{split}
B & = -\frac{\pi_B}{2} x^0+ P \,,
\\
I & = \ln \sinh|\sqrt{(A^2-9)\Lambda }x^0| \,,
\\
J & = \frac{\pi_J}{2m} x^0+ E \,.
\end{split}
\end{equation}
where $E$ and $P$ are new integration constants. This solution expressed in terms of the scale factor $a$ and the field $\delta$ is:
\begin{equation}
\label{roz}
\begin{split}
a & = \frac{1}{D^2 {(e^{ F x^0})}^2 \sinh ^ M |\sqrt{(A^2-9)\Lambda } x^0| } \,,
\\
\delta & = \frac{D^6 {(e^{ F x^0})}^6}{\sinh ^ W |\sqrt{(A^2-9)\Lambda } x^0|} \,,
\end{split}
\end{equation}
where
\begin{equation}
\label{const}
\begin{split}
D=e^{\frac{A}{9-A^2}E} \,,
\qquad
F=\frac{A}{9-A^2}\frac{\pi_J}{2m} \,,
\qquad
M=\frac{3-A^2}{9-A^2} \,,
\qquad
W=\frac{2A^2}{9-A^2} \,,
\end{split}
\end{equation}
and $A^2>9$.
It is also possible to relate variable $x^0$ with the proper time $\tau$  encountered by the comoving observers since both quantities are related by:
\begin{equation}
\label{vsl}
d\tau = \frac{\sqrt{-ds^2}}{c(x^0)} \,,
\end{equation}
where $ds^2 = - N^2 (dx^0)^2$. The relation \eqref{vsl} is a postulated generalisation of the general relativistic definition of the proper time  for the case of varying speed of light $c$ and it is assumed to be fulfilled in the cosmological frame. A justification for this particular generalisation of the proper time definition arises from the fact that we require light rays to follow null geodesics. The different way of introducing varying  $c$ into the metric was presented in ref.~\cite{GFEllis} where it was argued that  varying $c$ merely reflects the rescaling in the time coordinate and thus does not affect the value of the physical speed of light. In the definition \eqref{vsl}, however, $c$  does represent the physical value of the speed of light. Noticing that:
\begin{equation}
\label{laps}
N=\frac{1}{\sinh |\sqrt{(A^2-9)\Lambda } x^0|} \,,
\end{equation}
we obtain:
\begin{equation}
\label{conect}
\begin{split}
x^0 &= \frac{2}{\sqrt{(A^2-9)\Lambda}}  \arctanh \left(e^{\sqrt{(A^2-9)\Lambda}\bar{x}^0}\right) \,,
\qquad
\text{for $\bar{x}^0<0$} \,,
\\
x^0 &= \frac{2}{\sqrt{(A^2-9)\Lambda}}  \arctanh \left(e^{- \sqrt{(A^2-9)\Lambda}\bar{x}^0}\right) \,,
\qquad
\text{for $\bar{x}^0>0$} \,.
\end{split}
\end{equation}
where we have defined $d\bar{x}^0\equiv c(\bar{x}^0) d\tau$.
\begin{figure}[tbp]
\centering 
\includegraphics[width=.55\textwidth]{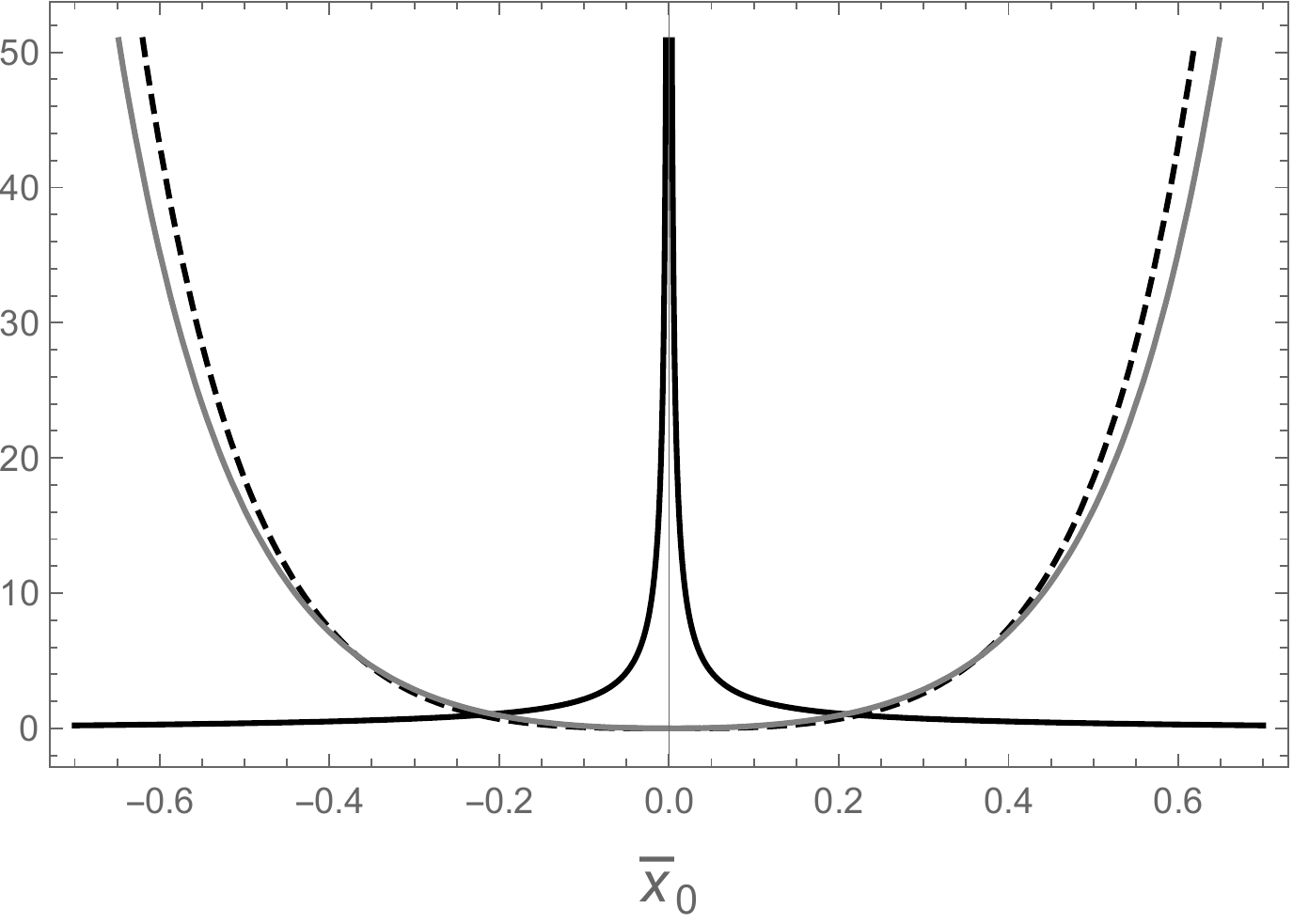}
\caption{\label{brancze} Qualitative behaviour of $a$ (gray), $c$ (black) and $G$ (dashed) before ($\bar{x}^0<0$) and after ($\bar{x}^0>0$) the big-bang singularity.}
\end{figure}
In Fig.\eqref{brancze} we have plotted the behaviour of the scale factor $a$ (gray), the speed of light $c$ (black), and the gravitational constant $G$ (dashed) for the pre-big-bang $\bar{x}^0<0$ and post-big-bang $\bar{x}^0>0$ phases. At $\bar{x}^0=0$ an interesting behaviour occurs:
as the universe approaches the singularity  ($a\rightarrow0$), the gravitational constant $G$ goes to zero while the speed of light $c$ reaches infinity.
This corresponds to the Newton's Mechanics limit represented by one of the vertex of the Bronshtein-Zelmanov-Okun cube~\cite{Duff,Okun}.
An analogous behavior is encountered in the case of the ekpyrotic~\cite{Khoury,Khoury2} and cyclic scenarios~\cite{Steinhardt,Steinhardt2}, where in the high curvature regime the universe,  after having undergone the phase of the accelerated contraction, faces the big-crunch/bang singularity and then bounces to enter the standard expanding phase. On the other hand,  we encounter a qualitatively different behavior in the pre-big bang scenario within the framework of string cosmology where the pre-big bang phase corresponds to expanding and accelerating universe approaching the strong coupling regime~\cite{Gasperini3}.

\section{Quantum mechanical behavior near the curvature singularity}
By examining the classical solution we find that the high-curvature (near big-bang) regime appears for $I\rightarrow \infty$, which is equivalent to $x^0\rightarrow \infty$ for $\bar{x}^0\rightarrow 0$, while the low-curvature (far away from big-bang) regime appears for $I\rightarrow -\infty$, which is equivalent to $x^0\rightarrow 0$. The latter regime corresponds to   $\bar{x}^0\rightarrow -\infty$ or to $\bar{x}^0\rightarrow \infty$ for the pre and post-big-bang branch, respectively. Both regimes are characterized by the following  asymptotic values of the momentum $\pi_I$:
\[ \pi_I = \left\{
  \begin{array}{l l}
    2 \tilde{V}_0\sqrt{\frac{\Lambda}{A^2-9}} & \quad \text{collapsing pre-big-bang }\\
    -2 \tilde{V}_0\sqrt{\frac{\Lambda}{A^2-9}} & \quad \text{expanding post-big-bang}
  \end{array} \right.\]
in the high-curvature limit and
  \[ \pi_I = \left\{
  \begin{array}{l l}
    2 \tilde{V}_0\sqrt{\frac{\Lambda}{A^2-9}} e^{-I} & \quad \text{collapsing pre-big-bang}\\
    -2 \tilde{V}_0\sqrt{\frac{\Lambda}{A^2-9}} e^{-I} & \quad \text{expanding post-big-bang}
  \end{array} \right.\]
  in the low-curvature limit.
In order to quantize the considered theory we make the following substitution:
\begin{equation}
\label{quant}
\begin{split}
\pi_J\rightarrow \hat{\pi}_J=-i \frac{\partial}{\partial J} \,,
\qquad
\pi_I\rightarrow \hat{\pi}_I=-i \frac{\partial}{\partial I} \,,
\qquad
\pi_B\rightarrow \hat{\pi}_B= -i \frac{\partial}{\partial B}\,.
\qquad
\end{split}
\end{equation}
The resulting Wheeler-De Witt equation is:
\begin{equation}
\label{WDW}
\left\{\frac{1}{4}\left[\frac{1}{m}\left(\frac{\partial^2}{\partial I^2}-\frac{\partial^2}{\partial J^2}\right)+ \frac{\partial^2}{\partial B^2}\right] -\bar{\Lambda}e^{-2I}\right\}\Phi=0 \,.
\end{equation}
It is interesting to point out that the sequence of the fields transformations given by \eqref{fred} and \eqref{fred2} reduces the problem of the factor ordering.
We look for the solution of eq. \eqref{WDW} which is separable~\cite{DabQ}:
\begin{equation}
\label{psi}
\Phi=\alpha(J)\gamma(B)\beta(I) \,,
\end{equation}
where
\begin{equation}
\begin{split}
\alpha(J) &= e^{i2k \sqrt{\frac{\tilde{V}_0}{A^2-9}}J} \,,
\\
\gamma(B) &= e^{i2lB} \,,
\\
\label{bessel}
\beta(I) &= J_{-i\pi^\infty_I}(\pi^\infty_I e^{-I}) \,.
\end{split}
\end{equation}
and
\begin{equation}
\begin{split}
\tilde{V}_0 \Lambda &=  k^2+l^2\,,
\\
\pi^\infty_I&= 2 \tilde{V}_0\sqrt{\frac{\Lambda}{A^2-9}} \,.
\end{split}
\end{equation}
The wave function $\Phi$, asymptotically for $I\rightarrow\infty$, is an eigenfunction of the momentum operator
$\hat{\pi}_I$ since:
\begin{equation}
\hat{\pi}_I \beta(I)=\pi^\infty_I\beta(I) \,.
\end{equation}
Thus, in the high-curvature regime  $\Phi$ contains only those modes which represent collapsing universe~\cite{DabKief,Gasperini2}.
In the low curvature limit  ($I \rightarrow -\infty$) the Bessel function $J_{in}$ can be decomposed in the following way:
\begin{equation}
J_{in}(z)=\Psi_1+\Psi_2 \,,
\end{equation}
where
\begin{equation}
\begin{split}
\Psi_1&=\sqrt{\frac{1}{2 \pi z}} e^{i\left(z-\frac{\pi}{2}in - \frac{\pi}{4}\right)} \,,
\\
\Psi_2&=\sqrt{\frac{1}{2 \pi z}} e^{-i\left(z-\frac{\pi}{2}in - \frac{\pi}{4}\right)} \,,
\end{split}
\end{equation}
and
\begin{subequations}\label{eq:y}
\begin{align}
\label{ev1}
\hat{\pi}_I \Psi_1 &= -z \Psi_1
\\
\label{ev2}
\hat{\pi}_I \Psi_2 &= z \Psi_2
\end{align}
\end{subequations}
where  $n\equiv-\pi^\infty_I$     and  $z\equiv\pi^\infty_I e^{-I}= 2 \tilde{V}_0\sqrt{\frac{\Lambda}{A^2-9}} e^{-I}$. Eqs. (\ref{ev1}) and (\ref{ev2}) therefore imply that $\Psi_2$ and $\Psi_1$ correspond,  to the pre- and post-big-bang branches  in their low curvature evolution phases, respectively. The probability of the transition form the pre-big-bang low curvature initial state to the post-big-bang low curvature final state is given by the reflection coefficient:
\begin{equation}
R=\frac{|\Psi_1|^2}{|\Psi_2|^2}=e^{2 \pi n}=e^{-4 \pi  \tilde{V}_0\sqrt{\frac{\Lambda}{A^2-9}}} \,.
\end{equation}
On the contrary, in the scenario provided by the string quantum cosmology~\cite{Gasperini,Gasperini2}, the initial state from which the quantum transition to the expanding phase occurs, represents an accelerating expanding universe.

\section{Conclusions}
We have considered a theory with  varying $c$ and $G$ where both constants are represented by non-minimally coupled scalar fields.  Such theory is equivalent to the Brans-Dicke theory. We  have examined the cosmological evolution near the curvature  singularity. It turned out that dynamics of the expansion can be reduced to the problem of one dimensional scattering of the point particle on the potential described with an exponential function. We have found that the cosmological evolution contains both the collapsing pre-big-bang phase and the expanding post-big-bang phase with the both  phases separated by the curvature singularity. At the curvature singularity $c$ reaches infinity while $G$ teds to zero. Thus, we have shown that the near curvature singularity evolution phase takes place well within the Newton's Mechanics regime given by $G = 0$ and $c = \infty$.
We have also examined the quantum mechanical behavior near the curvature singularity by  examining the Wheeler-De Witt equation. We have shown this to be equivalent to the quantum stationary scattering of the point particle on the exponential potential barrier. The solution of the Wheeler-De Witt in the low curvature regime is the superposition of the wave function that represents a particle with positive momentum moving towards the potential barrier and the wave function that represents a particle scattered on the barrier. The former correspond to a collapsing universe in the pre-big-bang phase while the latter represents an expanding universe in the post-big-bang phase. We have also calculated the probability of scattering, which is equivalent  to the probability of the transition from the pre-big-bang phase to post-big-bang phase.

\acknowledgments
I wish to thank Mariusz  D\c{a}browski for discussions. This project was financed by the Polish National Science Center Grant DEC-2012/06/A/ST2/00395.



\end{document}